\documentclass[aps,amsfonts,prl,twocolumn,showpacs]{revtex4-1}

\usepackage{pdfpages} 
\makeatletter
\AtBeginDocument{\let\LS@rot\@undefined}
\makeatother

\usepackage{subfigure}
\usepackage{textcomp}
\usepackage{graphicx}
\usepackage{bm}
\usepackage{amsmath,amssymb,amsthm}
\usepackage{amsfonts}
\usepackage{dsfont}
\usepackage[colorlinks=true,linkcolor=blue,urlcolor=blue,citecolor=blue]{hyperref}
\usepackage{multirow}
\usepackage{url}

\newcommand{\vpd}{\vphantom{\dagger}}
\newcommand{\vpp}{\vphantom{\prime}}

\newcommand{\FFA}[0]{\hat{F}^{\vpd}_{0}}

\newcommand{\DFFA}[0]{\hat{F} \left( \lambda \right)}
\newcommand{\hU}[1]{\hat{U}^{\vpd}_{#1}}
\newcommand{\ham}[0]{\hat{H}}
\newcommand{\com}[2]{[\, #1 \, , \, #2\, ]}

\newcommand{\ua}[0]{\uparrow^{\vpd}}
\newcommand{\da}[0]{\downarrow^{\vpd}}

\def\ket#1{\mathinner{|{#1}\rangle}}

\def\Ket#1{\left|#1\right>}

\def\beq{\begin{equation}}
\def\eeq{\end{equation}}
\def\bea{\begin{eqnarray}}
\def\eea{\end{eqnarray}}

\begin{document}

\title{Integrable many-body quantum Floquet-Thouless pumps}

\author{Aaron J. Friedman$^{1,2}$, Sarang Gopalakrishnan$^{3}$, and Romain Vasseur$^{2}$}
\affiliation{$^1$Department of Physics and Astronomy, University of California, Irvine, CA 92697, USA \\
$^2$Department of Physics, University of Massachusetts, Amherst, Massachusetts 01003, USA \\
$^3$Department of Physics and Astronomy, CUNY College of Staten Island, Staten Island, NY 10314;  Physics Program and Initiative for the Theoretical Sciences, The Graduate Center, CUNY, New York, NY 10016, USA}

\begin{abstract}

We construct an interacting integrable Floquet model featuring quasiparticle excitations with topologically nontrivial chiral dispersion. This model is a fully quantum generalization of an integrable classical cellular automaton. We write down and solve the Bethe equations for the generalized quantum model, and show that these take on a particularly simple form that allows for an exact solution: essentially, the quasiparticles behave like interacting hard rods. The generalized thermodynamics and hydrodynamics of this model follow directly. Although the model is interacting, its unusually simple structure allows us to construct operators that spread with no butterfly effect; this construction does not seem to be possible in other interacting integrable systems. This model illustrates the existence a new class of exactly solvable, interacting quantum systems specific to the Floquet setting. 

\end{abstract}

\maketitle

Periodically driven (or ``Floquet'') quantum systems have become an important and fruitful theme in condensed matter physics~\cite{RevModPhys.89.011004,doi:10.1080/00018732.2015.1055918,PhysRevLett.116.205301,0953-4075-49-1-013001,PhysRevX.4.031027,PhysRevB.82.235114,Lindner:2011aa,PhysRevX.3.031005,1367-2630-17-12-125014,PhysRevB.79.081406,Gorg:2018aa,PhysRevLett.116.125301,2019arXiv190501317H}: driving enables one to engineer and stabilize exotic states of matter, as has been experimentally demonstrated~\cite{gedik_floquet, floquet_amo_1, floquet_amo_2, floquet_amo_3}; moreover, driven systems can support phases such as anomalous insulators~\cite{Lindner:2011aa,PhysRevLett.107.216601,PhysRevX.3.031005,1367-2630-17-12-125014,PhysRevB.84.235108} and quantum time crystals~\cite{khemani_prl_2016,pi-spin-glass,else_floquet_2016,PhysRevX.7.011026,PhysRevLett.118.030401,PhysRevLett.119.010602,Zhang:2017aa,Choi:2017aa} that are absent in equilibrium. 
Driven free-particle systems are the best understood case: these have band structures that are compactified in both quasi-momentum and quasi-energy, and the fact that quasi-energy is only defined on a ring 
allows for new topological indices that are unrealizable for local lattice Hamiltonians~\cite{PhysRevB.79.081406,PhysRevB.82.235114,Lindner:2011aa,PhysRevLett.107.216601,PhysRevX.3.031005,1367-2630-17-12-125014,PhysRevB.84.235108}. For instance, Floquet systems can have a single chiral mode, which under Hamiltonian dynamics could only exist on the boundary of a higher-dimensional system~\cite{PhysRevB.82.235114,PhysRevX.3.031005,PhysRevLett.114.056801,PhysRevX.6.021013}. These topological indices are sharply defined for free-particle systems, and are potentially long-lived in some interacting lattice models~\cite{KUWAHARA201696,PhysRevLett.115.256803,PhysRevX.7.011018}; however, in general interactions heat a system up to infinite temperature, unless it is either integrable or many-body localized (MBL)~\cite{BAA,PhysRevB.75.155111,PalHuse,doi:10.1146/annurev-conmatphys-031214-014726,doi:10.1146/annurev-conmatphys-031214-014701,1742-5468-2016-6-064010,2018arXiv180411065A}. Although MBL can protect~\cite{HuseMBLQuantumOrder,BahriMBLSPT,BauerNayak} Floquet topological phases~\cite{khemani_prl_2016,PhysRevLett.114.140401, PhysRevB.93.245145,PhysRevB.93.245146,PhysRevB.93.201103,potter_classification_2016,PhysRevB.94.125105,PhysRevB.94.214203,PhysRevB.95.195128,PhysRevX.6.041070,PhysRevLett.120.150601}, 
these phases are \emph{localized} and do not host chiral modes. However, in addition to MBL systems, \emph{interacting} integrable systems are another broad class of systems---including the canonical Hubbard, Heisenberg, and Lieb-Liniger models---that do not heat up to infinite temperature~\cite{kinoshita, bloch_heisenberg, Rigol:2008kq}; whether distinctively Floquet versions of these models exist has been less discussed~\cite{SciPostPhys.2.3.021,10.21468/SciPostPhys.5.3.025,PhysRevLett.121.080401,PhysRevLett.121.030606}.

This work presents an interacting integrable Floquet model that has quasiparticles with nontrivial winding. This model is thus a many-body realization of a quantum Thouless pump~\cite{PhysRevB.27.6083,PhysRevB.82.235114,PhysRevX.7.011018}. 
Unlike previously proposed interacting integrable Floquet systems, our model is not smoothly connected to any Hamiltonian, and is thus an inherently Floquet model rather than an ``integrable Trotterization''~\cite{PhysRevLett.121.030606}. 
 This model is a fully quantum extension of an integrable cellular automaton (known as Rule 54, or the Floquet Fredrickson-Andersen (FFA) model~\cite{bobenko, prosen2016ffa, sg_ffa}) that has received much recent attention for its simplicity, which allows one to address explicitly various puzzles concerning  hydrodynamics and operator growth in generic interacting integrable systems~\cite{ prosen2016ffa, Prosen_2017,zakirov, sg_ffa, ghkv,klobas2018a,2019arXiv190100845B,2019arXiv190104521A}. The FFA model can be written as a Floquet unitary comprising local  
 quantum gates; however, it is classical in the sense that it maps computational-basis product states to one another. 
Although FFA has chiral quasiparticles, they do not disperse, but instead all have one of two group velocities, $\pm 1$. The dispersing FFA (DFFA) generalization we present here involves alternating the FFA dynamics with 
that of a particular strictly local Hamiltonian. This makes the model fully quantum by restoring 
 dispersion while preserving integrability. As we will show, our generalization also preserves enough of the simplicity of FFA 
 that the Bethe equations can be solved analytically---a remarkable feature for an interacting model. The reason this model is so simple is that the quantization for either species of quasiparticle depends only on the total number of quasiparticles of each species, and \emph{not} on their rapidities. This simplicity also manifests itself in the existence of special local operators that remain lightly entangled at all times, as in the FFA model~\cite{ghkv,2019arXiv190104521A}. This model is the first representative of a new class of interacting integrable models specific to the Floquet setting, featuring stable chiral quasiparticles.

\emph{Model}.---We consider a chain of $2L$ qubits (spins-$\frac{1}{2}$) with dynamics generated by the repeated application of the unitary evolution (Floquet) operator
\beq\label{model0}
\DFFA =   e^{-i \lambda \ham}  \prod\limits_{j \ \text{even}} \hU{j-1,j,j+1}  \prod\limits_{j \ \text{odd}} \hU{j-1,j,j+1},
\eeq
with gates $\hU{j-1,j,j+1} \equiv \, \mathrm{CNOT} (1 \rightarrow 2) \, \mathrm{CNOT} (3 \rightarrow 2) \, \mathrm{Toff.}(1,3 \rightarrow 2)$, in terms of controlled NOT (CNOT) and Toffoli gates~\cite{nielsen_chuang}; and $H$ is a Hamiltonian perturbation that we will specify below. In simpler terms, $\hU{j-1,j,j+1}$ is the instruction ``flip spin $j$ if one or both of its nearest neighbors is up.'' When $\lambda = 0$ this model reduces to the FFA model, $\hat{F} \left( 0 \right) = \FFA$.

\emph{FFA limit}.---On its own, $\FFA$ hosts two species of chiral quasiparticle excitations above the vacuum state $\ket{0} = \ket{\da \, \da \dots \da}$, indexed $\nu = +1$ for ``right movers'' and $\nu=-1$ for ``left-movers''. 
We regard the $2L$ physical sites as $L$ unit cells: the $n^{\rm th}$ unit cell contains the $A$ site $2n-1$ and $B$ site $2n$. If both of these sites are $\ua$, then there is a $\nu=+1$ right-moving doublon in cell $n$; if the $B$ site of cell $n-1$ and $A$ site of $n$ are both $\ua$, there is a $\nu=-1$ left-moving doublon in cell $n$. Additionally, we refer to isolated $\ua$'s as \emph{molecules}, which contain one of each mover: a molecule on the $A$ site of cell $n$ corresponds to both $\nu =\pm1$ movers in cell $n$; a $B$ molecule in cell $n$ corresponds to a $+$ at $n$ and $-$ at $n+1$. The molecule states $\da \ua \da$ arise during collisions between the two species.  Apart from these collisions, FFA acts by changing the positions (in unit cells) of the $\pm$ particles by $\pm 1$, and conserves independently the number of each, $N_{\pm}$.

Thus in the FFA model all quasiparticles on top of a given state have the same velocity. The structure of conservation laws in this model differs from that of generic interacting integrable models, in which a generalized Gibbs ensemble (GGE)~\cite{1742-5468-2016-6-064007} can be fully specified through the distribution of quasiparticle velocities. In the FFA model, there are only two velocities, which do not fully specify a state. The remaining conservation laws correspond to asymptotic ``spacings'' between adjacent quasiparticles of the same species~\cite{sg_ffa}. In the zero-density limit, the bare spacings between same-species quasiparticles are conserved, since all such quasiparticles have the same velocity. At nonzero densities, one can define an asymptotic spacing by accounting for interaction effects: e.g., suppose we have two $+$ quasiparticles that are $n$ steps apart; the quasiparticle on the right collides first with a $-$ quasiparticle and is time delayed by one step: therefore, while there is a $-$ quasiparticle between them, the two $+$ quasiparticles will be exactly $n - 1$ steps apart if their asymptotic spacing is $n$. Given a spin configuration, its asymptotic spacings can be found numerically by simulating its free expansion into vacuum~\cite{sg_ffa}.

\begin{figure}[t]
\includegraphics[width = 0.98\columnwidth]{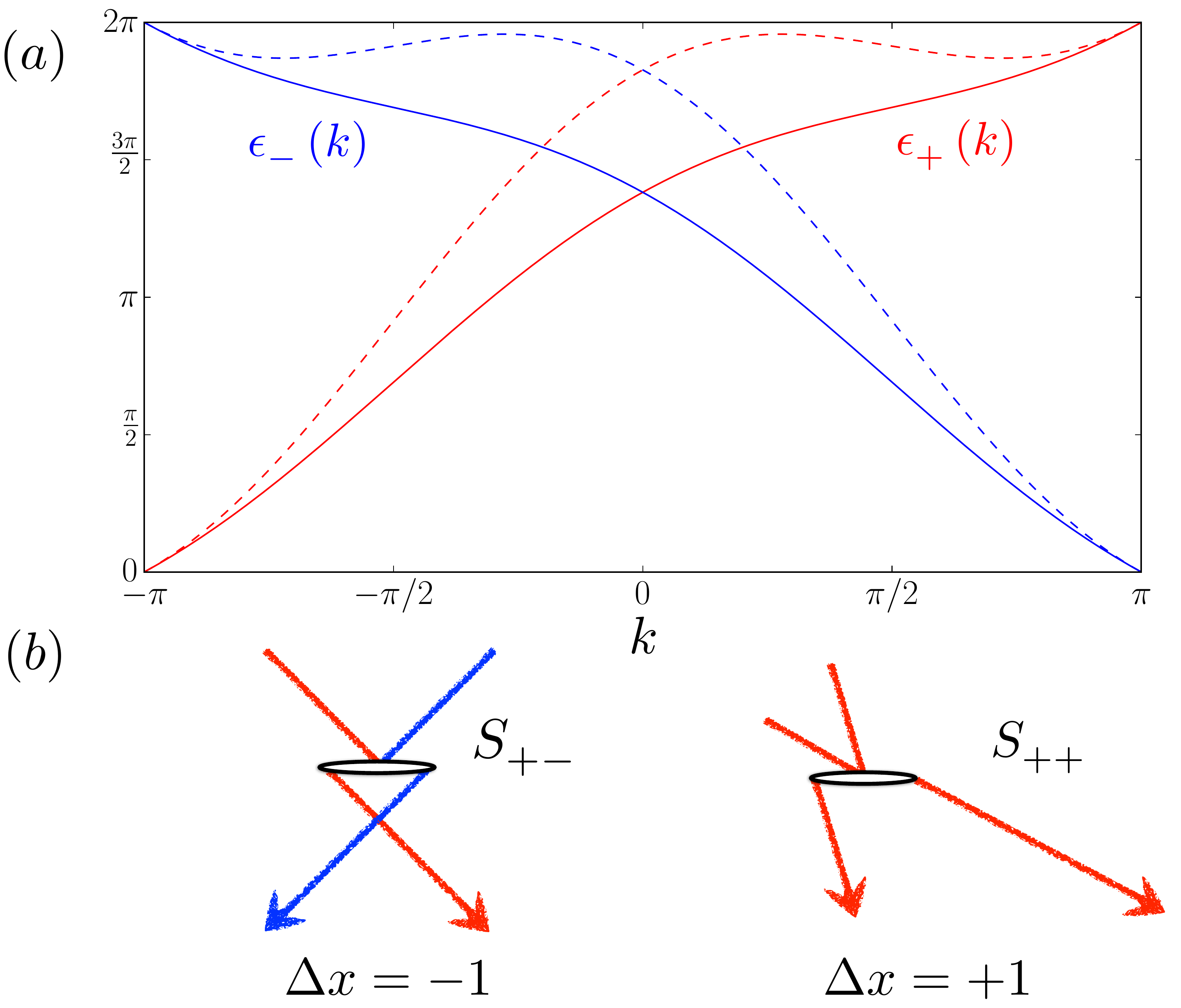} 
\caption{\label{fig1} {\bf Chiral quasiparticles in the DFFA model}. (a) Dispersion relations showing the (bare) single-particle Floquet quasi-energies $\varepsilon(k)$ for both $+$ and $-$ quasiparticles, for $\lambda=0.3$ (solid lines) and $\lambda=0.65$ (dashed lines). Both bands are topological (chiral) as they wrap around the periodic quasi-energy direction, and can only exist in a periodically driven system. Note that for $\lambda=0.65$, the $\pm$-particles can be left(right)-moving for some range of momenta. (b) Soliton gas picture. The scattering events in the DFFA model factorize onto simple two-body processes, which semi-classically correspond to a displacement $\Delta x =\pm 1$ after a collision, independently of the momenta of the quasiparticles.
}
\end{figure}

\emph{Adding dispersion}.---
We now construct $\ham$, the Hamiltonian part of \eqref{model0}, to generate dispersion while maintaining integrability. Conservation of particle number automatically precludes many simple terms, i.e. most single spin processes. 
Even a more complicated pair-hopping term like $\hat{\sigma}^+_i  \hat{\sigma}^+_{i+1}\hat{\sigma}^-_{i+2} \hat{\sigma}^-_{i+3}$ will not always conserve $N_{\pm}$: it can adjoin two doublons of the same species, producing another of the opposite type. 
The simplest $N_{\pm}$-conserving operator that disperses quasi-particles is $\hat{h}^{\vpd}_j \equiv \hat{d}^{\vpp}_{j-1} \hat{\sigma}^+_j \hat{\sigma}^+_{j+1} \hat{\sigma}^-_{j+2} \hat{\sigma}^-_{j+3} \hat{d}^{\vpp}_{j+4}$, where $\hat{d}^{\vpp}_j \equiv \frac{1}{2} (1 - \hat{\sigma}^z_j)$ [analogously, $\hat{u}^{\vpp}_j \equiv \frac{1}{2} (1 + \hat{\sigma}^z_j)$]. This term ``checks'' that it would not create any new quasiparticles before moving one.

Setting $\ham = \sum_j (\hat{h}^{\vpd}_j  + \hat{h}^{\dagger}_j )$ would give a simple dispersive extension of FFA; however, 
we cannot confirm that this preserves integrability. 
Hence, we add other terms: 
\bea\label{fullham}
\ham & = & \sum\nolimits_i \big( \hat{d}^{\vpp}_i \hat{\sigma}^+_{i+1} \hat{\sigma}^+_{i+2} \hat{\sigma}^-_{i+3} \hat{\sigma}^-_{i+4} \hat{d}^{\vpp}_{i+5} + \hat{d}^{\vpp}_i \hat{\sigma}^+_{i+1} \hat{\sigma}^-_{i+2} \hat{d}^{\vpp}_{i+3} \notag \\
& & \qquad + \hat{d}^{\vpp}_i  \hat{\sigma}^+_{i+1}  \hat{\sigma}^+_{i+2} \hat{u}^{\vpp}_{i+3} \hat{d}^{\vpp}_{i+4} + \mathrm{refl.} \nonumber \\
&& \qquad + \hat{d}^{\vpp}_i \hat{\sigma}^+_{i+1} \hat{\sigma}^+_{i+2} \hat{\sigma}^-_{i+3} \hat{u}^{\vpp}_{i+4} \hat{u}^{\vpp}_{i+5} + \mathrm{refl.} \nonumber \\
&& \qquad + \hat{d}^{\vpp}_i \hat{u}^{\vpp}_{i + 1} \hat{\sigma}^+_{i+2} \hat{u}^{\vpp}_{i+3} \hat{u}^{\vpp}_{i+4} + \mathrm{refl.} \nonumber \\
& & \qquad + \hat{u}^{\vpp}_i \hat{u}^{\vpp}_{i+1} \hat{\sigma}^+_{i+2} \hat{\sigma}^-_{i+3} \hat{u}^{\vpp}_{i+4} \hat{u}^{\vpp}_{i+5} \big) + \mathrm{h.c.},
\eea
where ``refl.'' indicates that one should reverse the sequence of indices in the previous term. In the quasiparticle language, $\ham$~\eqref{fullham} maps a configuration $\sigma$ to a uniform superposition of all configurations $\sigma^{\prime}$ with a single quasiparticle moved by one unit cell, provided $N_{\pm}$ are preserved. Although this is precisely what one expects of a generic dispersing term, in this system it requires multiple microscopic processes. 

We remark that \eqref{fullham} commutes with the FFA unitary, $\FFA$. It acts nontrivially, regardless, because $\FFA$ has exponentially degenerate eigenstates: for a given $N_{\pm}$ in a system of size $L$, there are only $O(L^2)$ eigenvalues, but exponentially many basis states, corresponding to different quasiparticle positions. 
The perturbation~\eqref{fullham} lifts the degeneracy in this subspace, and thus makes the dynamics fully quantum. This perturbation cures many pathological features of the FFA model that are due to these degeneracies, such as its failure to equilibrate to the diagonal ensemble~\cite{sg_ffa}.

\emph{Single-quasiparticle sectors}.---We start by finding the eigenstates of \eqref{model0} for a single $\pm$ quasiparticle, $\Ket{j,\pm} = \sigma^x_{2j} \sigma^x_{2j \pm 1-1} \Ket{0}$. The Fourier transform is an eigenstate of $\DFFA$,
\beq\label{eqDispersion}
\DFFA \Ket{k, \nu} = {\rm e}^{-i \nu k - 2 i \lambda \cos k} \Ket{ k, \nu } ,
\eeq
where $\nu = \pm 1$. 
Here, $\lambda$ controls the strength of the dispersing term, and $k = 2\pi m/L$ for integer $m$, with $L$ the system size 
in unit cells. This model thus has two chiral bands (Fig.~\ref{fig1}). For $\lambda < 1/2$, all $+$ ($-$) quasiparticles have right- (left-) moving group velocities, but for $\lambda > 1/2$, both species have left- and right-moving quasiparticles. The group velocities of $\pm$ quasiparticles are given by $v^0_{\pm, k} = \pm 1 - 2 \lambda \sin k$. These chiral bands are characterized by a quantized winding number $\nu = \int_{-\pi}^\pi \frac{dk}{2 \pi} v_{\pm,k}^0 = \pm 1$, which is the topological invariant characterizing Thouless' quantized charge pump~\cite{PhysRevB.27.6083,PhysRevB.82.235114,PhysRevX.7.011018}.

\emph{Bethe Ansatz solution}.---We now move on to multi-particle sectors. We note, first, that in the absence of left-movers, the FFA evolution is just a trivial global translation. In this purely right-moving sector, the dynamics of $+$ quasiparticles consists of hopping and hardcore nearest-neighbor repulsion. 
The scattering phase shift between particles of the same species is thus $S_{\rm ++}(k_2 ,k_1) = S_{\rm --}(k_2 ,k_1) = \mathcal{S} (k_2 ,k_1)= - {\rm e}^{i (k_2 - k_1)}$.
Meanwhile, the scattering between left and right movers is engineered to retain the FFA form such that the phase shift after a collision is $S_{-+} ( k_+, k_- ) = \tilde{\mathcal{S}} ( k_+, k_- ) = +e^{i \left( k_+ - k_- \right) }$, and no meaning is ascribed to the order of the arguments. 
Higher-body collisions factorize onto the two-body scattering processes, ensuring integrability. For a many-body state with fixed $(N_+,N_-)$,  where $\lbrace k^{\pm}_j \rbrace$ refer to the momenta of the $\pm$-quasiparticles, we find the following quantization condition~\cite{suppmat}
\bea
e^{i k^+_j L} &= \prod\limits_{\substack{n=1 \\ n \neq j}}^{N_+} \mathcal{S} \left( k^+_j \, , \, k^+_n \right) \, \prod\limits_{m=1}^{N_-} \tilde{\mathcal{S}}^{*} \left( k^+_j \, , \, k^-_m \right), \notag \\
e^{i k^-_j L} &= \prod\limits_{\substack{n=1 \\ n \neq j}}^{N_-}  \mathcal{S} \left( k^-_j \, , \, k^-_n \right) \, \prod\limits_{m=1}^{N_+} \tilde{ \mathcal{S}} \left( k^-_j \, , \, k^+_m \right). \label{baeq}
\eea
These quantization conditions have the same form as Bethe equations familiar in Hamiltonian systems. Translational invariance and the recurrence properties of the FFA model (with which the Hamiltonian~\eqref{fullham} commutes) impose two further constraints. First, we require that $\sum_j k_j^+ + \sum_j k_j^- = K$, where $K = 2\pi m/L$ with $m$ an integer is one of the allowed global momenta. Second, we require that the relative momentum $\sum_j k_j^+ - \sum_j k_j^- = \Theta$, where 
\beq
\Theta = \frac{2\pi N_\theta + (N_+ - N_- - L ) K}{L + N_- + N_+},
\eeq
with $ 1 \leq N_\theta \leq (L + N_- + N_+)$ an integer, unless $L + N_- + N_+$ is even, in which case $N_{\theta}$ must be as well~\cite{suppmat}. Finally, no two quasiparticles of the same species may occupy the same momentum state.  With these constraints the solutions~\eqref{baeq} fully characterize the eigenstates in a finite system, and the corresponding quasi-energy ${\rm e}^{-i \varepsilon}$ of the Floquet unitary~\eqref{model0} reads  $\varepsilon = \sum_{\nu = \pm} \sum_{n=1}^{N_{\nu} } ( \nu k^{\nu}_n + 2 \lambda \cos k_n^{\nu})$. 

Remarkably, these equations are simple enough that they can be solved exactly for any finite system. The set of allowed momenta for either species, $\nu$ is
\beq\label{quantization}
k_j^{\nu}  =  \frac{\pi (2 m^{\nu}_j + N_{\nu}  -1 ) - \nu \Theta}{L - N_{\nu} + N_{\bar{\nu}}}, 
\eeq
with $\bar{\nu} = - \nu$ and $1 \leq m^{\nu}_j  \leq  L - N_{\nu} + N_{\bar{\nu}}$. The number of available $m^{\pm}_j$ decreases with the total number of $\pm$ movers 
because neighboring unit cells cannot both host $\pm$'s without a $\mp$ between them. 
We also note that the quantization condition depends on the total number and momentum of the $\pm$ quasiparticles, not on the details of their distribution.
Relatedly, \eqref{baeq} and \eqref{quantization} do not depend on $\lambda$, and thus also apply to $\FFA$, though in that model, the phase shift between 
quasiparticles of the same species is ill-defined as they move in unison and never collide.

\begin{figure}[t]
 {
\includegraphics[width = 0.98\columnwidth]{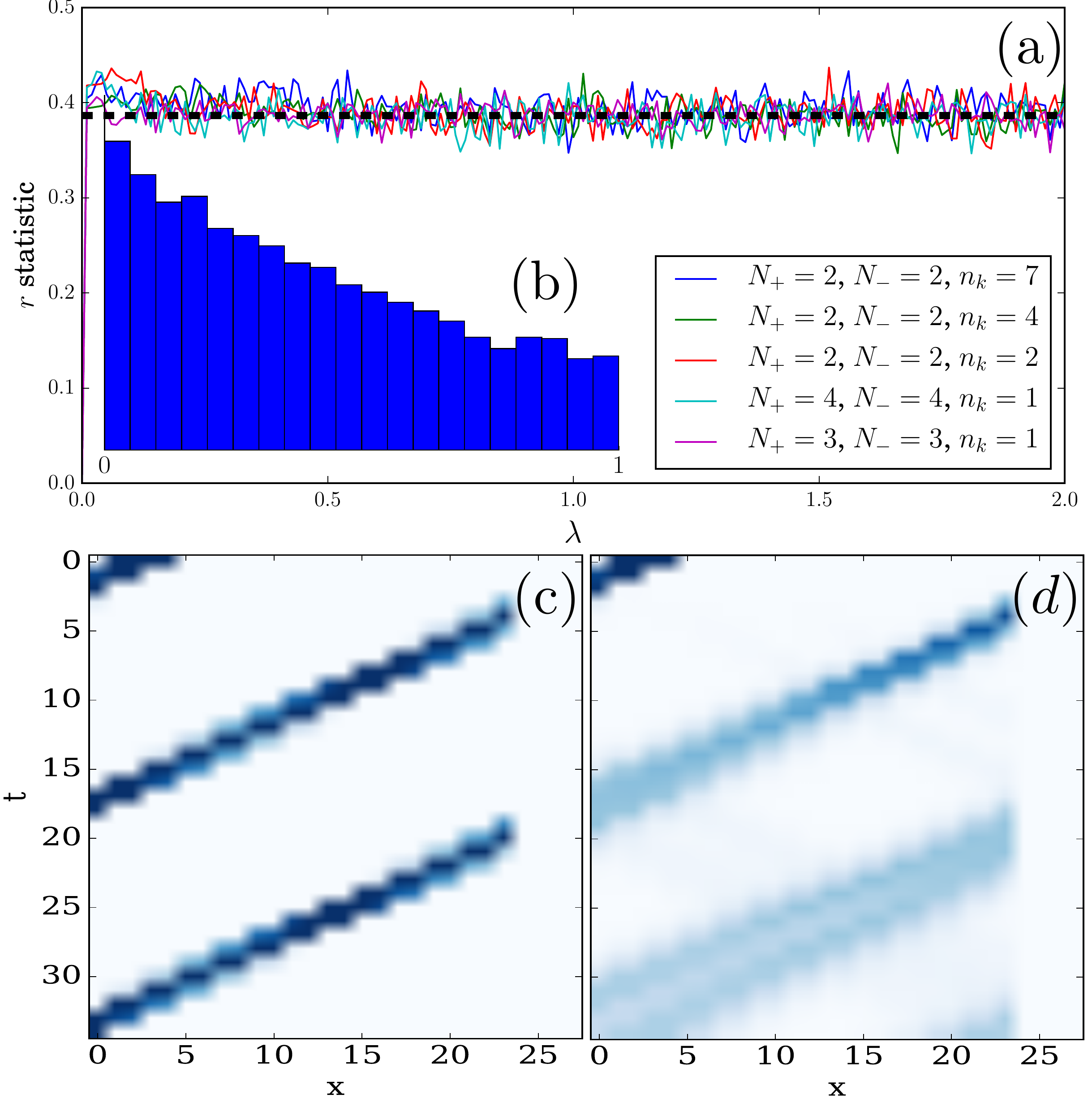} 
\caption{\label{fig2} {\bf Numerical results}. Quasi-energy level statistics for several values of $N_{\pm}$ and $K$ at $L=9$: (a) the $r$ ratio shows good agreement with a Poisson distribution (dashed line) for all $\lambda >0$; (b) the distribution of $r$ for $\lambda = 1.0$ (inset) does not show level repulsion, consistent with integrability. Plot of the OTOC $C(t)$ for $L=14$ unit cells with $N_+=1$ and $N_-=2$, for (c) $\lambda = 0$, corresponding to FFA, and (d) $\lambda=0.05$, where we see that the OTOC does not ``fill in'' behind the front except through the dispersion of the perturbed quasiparticle. }
}
\end{figure}

The DFFA model corrects several pathological features of FFA. This can be seen numerically by analyzing the (quasi-)energy level statistics, which does not show level repulsion (Fig.~\ref{fig2}), consistent with integrability. We also checked that the value of the ``$r$-ratio''~\cite{PhysRevB.75.155111} is consistent with a Poisson distribution for all $\lambda >0$. 

\emph{Thermodynamics}.---
 In the thermodynamic limit, one defines densities of quasiparticles at a given species and rapidity, $\rho_{\pm} (k)$, as well as total densities of states $\rho_{\pm}^{\rm tot} (k)$, related via the Bethe equations
\bea\label{tba0}
2\pi \rho_{\pm}^{\rm tot}(q)  = & 1 + n_{\mp} - n_{\pm}, 
\eea
where  $n_{\pm} \equiv \int_{-\pi}^\pi dq \rho_{\pm}(q) = N_{\pm}/L$. These equations follow from the continuum limit of~\eqref{baeq}, with the scattering kernels ${\cal K}_{\nu \nu'} = \frac{1}{2 \pi i} \frac{d }{d k}\, \ln S_{\nu \nu'}$ with $\nu, \nu' \in \left\{ +,-\right\}$ given by $\mathcal{K}_{++} = \mathcal{K}_{--} = 1/(2\pi)$, $\mathcal{K}_{+-} = \mathcal{K}_{-+} = -1/(2\pi)$. 

Starting with these equations, one can straightforwardly construct generalized equilibrium states of this Floquet system. We emphasize that since the DFFA model is integrable, its dynamics lead to non-trivial steady states that are distinct from featureless infinite temperature states that would be expected for generic interacting Floquet systems.  For concreteness we focus on generalized equilibrium states characterized by a given density of $\pm$ quasiparticles via the partition function $\mathcal{Z} = \sum_{\{\sigma\}} e^{-\mu_- N_- - \mu_+ N_+}$, but our discussion extends naturally to arbitrary GGEs for this model. In terms of quasiparticle densities, the partition function reads $\mathcal{Z} \sim \int {\cal D} \rho^{\vpp} e^{L \int d k S_{\rm YY}} e^{ - L \mu_+ \int dk \rho^{\vpp}_{k,+} - L \mu_- \int dk \rho^{\vpp}_{k,-} } $ where $S_{\rm YY}$ 
is the so-called Yang-Yang entropy~\cite{YY,Takahashi} associated with the occupation of quasiparticle states. In the thermodynamic limit $L \to \infty$, these integrals are dominated by their saddle point, giving rise to thermodynamic Bethe Ansatz (TBA) equations in a manner entirely analogous to Hamiltonian integrable systems where energy is conserved~\cite{Takahashi}. This leads to the following equations for the occupation numbers (Fermi factors) $\theta_{\nu}(k) \equiv \rho_{\nu}(k)/\rho^{\rm tot}_{\nu} (k) \equiv (1 + e^{\epsilon_{\nu} (q)})^{-1}$ which turn out to be independent of $k$:
\beq\label{tba1}
\epsilon_\pm = \mu_\pm + \log \left( \frac{1 + e^{-\epsilon_\pm}}{1 + e^{-\epsilon_\mp}} \right).
\eeq
Together with \eqref{tba0} this forms a complete characterization of the generalized Gibbs ensemble.  For $\lambda = 0$ (FFA model), the properties of this ensemble can also be derived by a transfer-matrix calculation~\cite{ghkv}; these approaches give equivalent results~\cite{suppmat}. 

\emph{Hydrodynamics and soliton gas}.--- The evolution from local to global equilibrium in the DFFA model can be described using the recently developed theory of generalized hydrodynamics (GHD)~\cite{Doyon, Fagotti} -- see also~\cite{SciPostPhys.2.2.014,GHDII, PhysRevLett.119.020602, BBH, solitongases, piroli2017, PhysRevB.96.081118,PhysRevB.97.081111,alba2017entanglement,PhysRevB.96.020403, bertini2018low, BBH0,1751-8121-50-43-435203, PhysRevLett.119.195301, 2017arXiv171100873C, PhysRevLett.120.164101,doyon2018exact,PhysRevLett.122.090601}. This is equivalent to treating this quantum system semiclassically as a gas of solitons~\cite{solitongases}. There are two species of solitons $\pm$ whose bare velocities are given by the dispersion relation~\eqref{eqDispersion}, so that $v^0_{\nu,k} = \nu - 2 \lambda \sin k$ with $\nu =\pm$. When solitons collide, they interact {\it via} a $k$-dependent phase-shift which leads to a semi-classical displacement $\Delta x = 2 \pi {\cal K}$ (Wigner time delay); i.e. $\Delta x = 1$ if the two quasiparticles are of the same species, and $\Delta x = -1$ otherwise (Fig.~\ref{fig1}). To leading order (Euler hydrodynamics), this leads to a dressing of the velocities due to collisions~\cite{PhysRevLett.113.187203,Doyon, Fagotti}, with the effective velocities in a state with quasiparticle densities $\rho_{\pm,k}$ given by 
\beq
v_{\pm,k} = v_{\pm,k}^0 + \int dq (v_{\pm,k }- v_{\pm,q }) \rho_{\pm,q} - \int dq (v_{\pm,k }- v_{\mp,q }) \rho_{\mp,q}.
\eeq
Each species of solitons thus behaves as a simple one-dimensional classical hard rod gas with unit length, while collisions between $+$ and $-$ solitons correspond formally to hard rods of negative length. Diffusive corrections to this ballistic picture follow from recent GHD results~\cite{dbd1, ghkv, dbd2,gv_superdiffusion, denardis_superdiffusion}, and are especially simple for DFFA~\cite{suppmat}. 

\emph{Operator dynamics}.---The rapidity-independent scattering kernels in the DFFA model have important consequences for operator spreading, which is simpler here than in generic integrable models~\cite{ghkv}. In the generic case, any operator creates a ``butterfly cone'' that fills in at late times: a spatially local operator has a spread of momenta and thus of group velocities, and the velocity-dependence of the scattering kernel implies that perturbing the velocity of one quasiparticle will affect the trajectories of all the others. This does not happen either in the hard rod gas or in the DFFA model, since the scattering kernel in these models is velocity-independent and consequently, any perturbation that preserves $N_\pm$ will only affect the state of one quasiparticle. Thus the butterfly cone, measured via the out-of-time-order correlator~\cite{lo_otoc, maldacena2016bound, motrunich2018, 2018arXiv180305902K} (OTOC) $ C \left( x , t \right) \equiv \frac{1}{2} \left|  {\rm Tr } \left\{ \com{\hat{h}^{\vpd}_{j=2}}{\hat{\sigma}^z_x \left( t\right)}^2 \right\} \right| $ does not ``fill in'' except through the dispersion of the perturbed quasiparticle (Fig. \ref{fig2}). 
The existence of such operators with simple matrix elements has to do with the structure of the Bethe ansatz equations: in a generic integrable model, changing the rapidity of one quasiparticle would alter the quantization condition for all the others. Therefore, the matrix element from a reference state to a state with one shifted quasiparticle would be suppressed by overlap factors from all the other quasiparticles that have their momenta slightly shifted. In the DFFA model, by contrast, this quantization condition depends only on a few aggregate properties of the quasiparticle distribution, so the matrix element for changing the state of a single quasiparticle is not parametrically suppressed. 

\emph{Conclusion}.--- In summary, we present and solve exactly a Floquet model that is the first of its kind in a number of respects. It is the first example of an interacting integrable Floquet model that is not smoothly deformable to Hamiltonian dynamics (integrable Trotterizations~\cite{PhysRevLett.121.030606}), and also not classically simulable (FFA). In fact, our solution of the dispersing model has provided insight into the physics of FFA, which prior to this work was not confirmed to be integrable in the Yang-Baxter sense; and the dispersing model regularizes several pathological features of FFA while making the model more reminiscent of typical quantum systems. 
Despite the complicated nature of the Hamiltonian terms, the resulting Bethe~\eqref{baeq} and TBA equations~\eqref{tba1} are the simplest of any interacting integrable model as far as we are aware. This model shows the existence of interacting Floquet models with stable chiral quasiparticles, and suggests a route to finding others, building on integrable cellular automata~\cite{bobenko, ica1, ica2, zakirov}; it would be interesting to find other examples in the future.

\emph{Acknowledgements}.--- We thank D. Huse, V. Khemani, A. Nahum, T. Prosen and B. Ware for illuminating discussions. This research was supported by the National Science Foundation via Grants DGE-1321846 (Graduate Research Fellowship Program, AJF), DMR-1455366 (AJF) and DMR-1653271 (SG); and the US Department of Energy, Office of Science, Basic Energy Sciences, under Award No. DE-SC0019168 (RV).

\bibliography{refs}

\bigskip

\onecolumngrid
\newpage



\section*{Supplementary material}

# Supplemental Material for "Integrable many-body quantum Floquet-Thouless pumps"

Aaron J. Friedman,[1,2] Sarang Gopalakrishnan,[3] and Romain Vasseur[2]

[1]*Department of Physics and Astronomy, University of California, Irvine, CA 92697, USA*
[2]*Department of Physics, University of Massachusetts, Amherst, Massachusetts 01003, USA*
[3]*Department of Physics and Astronomy, CUNY College of Staten Island,
Staten Island, NY 10314; Physics Program and Initiative for the Theoretical Sciences,
The Graduate Center, CUNY, New York, NY 10016, USA*
(Dated: May 7, 2019)

## I. THE NON-DISPERSING MODEL

The model we study in the main text is a modification of the Floquet Fredrickson-Andersen (FFA) model, an integrable cellular automaton (also known as Rule 54). Our model is a two step Floquet evolution operator, consisting of the bare FFA unitary operator, $\hat{F}_0$, as well as a "Hamiltonian" evolution term $e^{-i\lambda \hat{H}}$. In principle, $\hat{F}_0$ is applied first, however by construction $[\hat{H}, \hat{F}_0] = 0$, and thus the Hamiltonian and FFA have simultaneous eigenstates.

The FFA unitary $\hat{F}_0$ is defined for spins half on the sites of a one dimensional lattice – numbered $1, \ldots, 2L$ for convenience – and comprises a two step cycle: first, every odd spin is flipped unless both neighboring spins are down (in the $\sigma_j^z$ basis); the process is then repeated for the even spins. In the spin half language, we can write

$$\hat{F}_0 = \hat{W}_{\text{even}} \cdot \hat{W}_{\text{odd}}, \quad \text{with} \quad \hat{W}_{\text{sites}} \equiv \bigotimes_{j \in \text{sites}} \left[ \sigma_j^x \left( \mathbb{1} - \hat{d}_{j-1} \hat{d}_{j+1} \right) + \hat{d}_{j-1} \hat{d}_{j+1} \right], \tag{1}$$

where $\hat{d}_j = \frac{1}{2}(\mathbb{1}_j - \sigma_j^z)$ is the projector onto $\downarrow$ on site $j$ in the $\sigma^z$ basis. Analogously, we define the projector onto $\uparrow$ as $\hat{u}_j = \frac{1}{2}(\mathbb{1}_j + \sigma_j^z)$ for later use. We can also write (1) in terms of logic operators[1]: $\hat{W}_{\text{sites}} \equiv \bigotimes_{j \in \text{sites}} \hat{U}_{j-1,j,j+1}$ where $\hat{U}_{j-1,j,j+1} \equiv \text{CNOT}(1 \rightarrow 2)\text{CNOT}(3 \rightarrow 2)\text{Toff.}(1,3 \rightarrow 2)$, in terms of controlled NOT (CNOT) and Toffoli gates.

This model has a "vacuum" $|0\rangle = |\downarrow\downarrow \ldots \downarrow\downarrow\rangle$ upon which $\hat{F}_0$ acts as the identity. The elementary excitations are "doublons", or pairs of neighboring flipped spins $\uparrow\uparrow$, and come in two flavors, depending on whether the first spin is on an even or odd site. Thus, we regard the lattice of consisting of $L$ two-site unit cells, labelled $A$ and $B$, corresponding respectively to odd and even sites in the original enumeration. From the vacuum, we create a $\nu = -$ doublon in the $n^{\text{th}}$ unit cell by flipping the sites $2n-2$ and $2n-1$, or a $\nu = +1$ doublon in the $n^{\text{th}}$ unit cell by flipping sites $2n-1$ and $2n$. In isolation, $\hat{F}_0$ will act by moving the $\nu = -1$ doublons one unit cell to the left, and the $\nu = +1$ doublons one unit cell to the right, and hence we refer to these respectively as left- and right-movers, or more commonly, by their displacement $\delta x = \pm 1$ under FFA. FFA conserves the total momentum, as well as the respective numbers of left and right movers[2].

The single particle excitations of this model are given by plane wave superpositions of these excitations:

$$|k, \nu\rangle = L^{-1/2} \sum_{n=0}^{L} e^{ikn} \sigma_{2n-1}^x \sigma_{2n+\nu-1}^x |0\rangle \quad \text{s.t.} \quad \hat{F}_0 |k, \nu\rangle = e^{-i\epsilon_\nu(k)} |k, \nu\rangle, \quad \text{with} \quad \epsilon_\nu(k) = \nu k, \tag{2}$$

where $k$ is the momentum. Note that the unit cell translation operator $\hat{T}$ commutes with both $\hat{F}_0$ and $\hat{H}$; its eigenvalues are $e^{-ik}$ for $k = 2\pi q/L$ for integer $q \in \{1, 2, \ldots, L\}$. From the above expression, we see that $\hat{F}_0$ has the same action as $\hat{T}$ on right movers and $\hat{T}^{-1} = \hat{T}^\dagger$ on left-movers. Because of the chiral nature of FFA and its quasi-particles, this property will be true in any sector with only one of the two particle species present.

To describe sectors with additional movers, we must understand the counting of the number of movers in a given configuration. When one of each movers is present, at some point $\hat{F}_0$ will cause them to collide, realizing one of two "molecule" states, which are special configurations with a single, isolated up spin. One obtains an $A$ or $B$ molecule in the $n^{\text{th}}$ unit cell when, respectively, the spin on the $A$ or $B$ site of that unit cell is up, with both of its neighbors down ($\downarrow_{2n-2} \uparrow_{2n-1} \downarrow_{2n}$ or $\downarrow_{2n-1} \uparrow_{2n} \downarrow_{2n+1}$). In the former case ($A$), both the left- and right-mover are taken to be in unit cell $n$ (i.e. they have the same position), in the latter, the right mover is still at cell $n$, but the left-mover is in cell $n+1$.

Thus, we say there is a $\nu = +1$ right-mover in unit cell $n$ if both spins in the unit cell are up (in which case the mover is a doublon; this is independent of the neighboring spins), or if *either* spin in the unit cell is up, with both neighbors down. The condition for a $\nu = -1$ left-mover is *shifted to the left by one physical spin compared to the right*: if the $B$ site of cell $n-1$ and the $A$ site of cell $n$ are up, there is a left-moving doublon in cell $n$ (again, independent of neighboring spins); if *either* of these spins is up and both its neighbors down, then there is a left-mover in site $n$

that is part of a molecule. With this, rather than label states by the configurations of the physical spin half degrees of freedom, we will do so by the positions of the various movers and their form (i.e. doublons vs. molecules).

Looking to more-body sectors, if only one species is present, $\hat{F}_0$ acts as a translation operator (or its inverse, depending on $\nu = \pm 1$). There are never collisions between particles of the same species under FFA, only those of opposite species. Unlike a conventional translation operator, FFA is special in that it has its own $S$ matrix associated with this scattering of left and right movers, which we will observe in sectors containing both species $\nu$.

Let us now consider $\hat{F}_0$ when $N_+ = N_- = 1$, which will provide insight into generic many-body sectors. Acting on the majority of configurations, $\hat{F}_0$ moves the left and right movers one site in their namesake direction, however when the two are nearby, we obtain the following collisions in the $|x_+, x_-\rangle$ basis for doublons and $|\alpha, x\rangle$ basis for molecules, with $\alpha = A, B$:

$$\begin{array}{ccccccc}
\downarrow\uparrow\uparrow\downarrow\!\!\stackrel{*}{\downarrow}\!\!\downarrow\uparrow\uparrow\downarrow & \longrightarrow & \downarrow\uparrow\stackrel{*}{\uparrow}\uparrow\downarrow & \longrightarrow & \downarrow\stackrel{*}{\uparrow}\downarrow & \longrightarrow & \downarrow\uparrow\uparrow\stackrel{*}{\downarrow}\uparrow\uparrow\downarrow \\
{}_+ \quad {}_- & & {}_+ \quad {}_- & & B & & {}_- \quad {}_+ \\
|n-1, n+2\rangle & \longrightarrow & |n, n+1\rangle & \longrightarrow & |B, n\rangle & \longrightarrow & |n+1, n\rangle
\end{array} \quad (3)$$

$$\begin{array}{ccccccc}
\downarrow\uparrow\uparrow\stackrel{*}{\downarrow}\uparrow\uparrow\downarrow & \longrightarrow & \downarrow\stackrel{*}{\uparrow}\downarrow & \longrightarrow & \downarrow\uparrow\stackrel{*}{\uparrow}\uparrow\downarrow & \longrightarrow & \downarrow\uparrow\uparrow\downarrow\!\!\stackrel{*}{\downarrow}\!\!\uparrow\uparrow\downarrow \\
{}_+ \quad {}_- & & A & & {}_- \quad {}_+ & & {}_- \quad\quad {}_+ \\
|n-1, n+1\rangle & \longrightarrow & |A, n\rangle & \longrightarrow & |n, n\rangle & \longrightarrow & |n+1, n-1\rangle
\end{array}, \quad (4)$$

where in the top line the $*$ site is $2n$ (3), and in the bottom it's $2n-1$ (4). As in the single-particle (or single-species) sectors, $\hat{F}_0$ acts by cycling through the states in a closed "orbit", much like the eigenstates of the translation operator, $\hat{T}$.

For $N_+ = N_- = 1$, $\hat{F}_0$ changes the separation of the movers $\delta \equiv x_- - x_+$ by two (excepting a "delay" step which leaves both in place) (3–4). Hence if $L$ is even, $\hat{F}_0$ preserves $\delta$ modulo 2, yielding two distinct orbits with degenerate eigenvalues under FFA, distinguished by $\delta \bmod 2$; if $L$ is odd, then both types of collisions must occur before we return to the initial configuration. Much of this discussion will apply to sectors with more movers, and the general method for constructing eigenstates will be the same.

Let us now construct eigenstates of $\hat{F}_0$ for $N_+ = N_- = 1$ and even $L = 2\ell$. As noted, there are two degenerate orbits of $\ell + 1$ states under $\hat{F}_0$, corresponding respectively to the collision processes (4) and (3), or equivalently, $A$ vs. $B$ molecules, or $\delta$ even vs. odd. The fact that the orbits are closed follows from the identity $\hat{F}_0^{\ell+1}\hat{T}^{\pm\ell} = \mathbb{1}$, which immediately dictates the eigenvalues of $\hat{F}_0$,

$$\hat{F}_0|n, k, \alpha\rangle = e^{-i\theta_{n,k}}|n, k, \alpha\rangle, \quad \theta_{n,k} = \frac{2\pi n - \ell k}{\ell + 1} = \frac{4\pi n - Lk}{L + 2}, \quad (5)$$

where $\alpha = A, B$ designates the orbit based on its molecule, as depicted in (3–4). The eigenvalue $\theta_{n,k} = \sum_\pm \pm k_\pm$ is the *relative momentum* and $k = \sum_\pm k_\pm$ is the total momentum, set by translation invariance as usual.

We now formulate the eigenstates of $\hat{F}_0$ explicitly: we will first define a natural basis for this sector, which will allow us to conveniently express the allowed orbits, and finally, massage these eigenstates into the traditional plane-wave form common to integrable systems. The first step is the definition of a translation-invariant basis for this sector, indexed by the separation $\delta$

$$|\delta, k\rangle = L^{-1/2}\sum_{j=1}^{L} e^{ikj}|x_+ = j, x_- = j + \delta \bmod L\rangle \quad (6)$$

for the doublon configurations, and for the molecules $\alpha = A, B$:

$$|\alpha, k\rangle = L^{-1/2}\sum_{j=1}^{L} e^{ikj}|\alpha, j\rangle, \quad (7)$$

though in this sector, one can also interpret the molecules as additional values of $\delta$. We define the eigenstates of $\hat{F}_0$ with respect to some "reference configuration", which we choose to be the corresponding molecule states $|\alpha, k\rangle$ for notational convenience, i.e. the eigenstates are formed as

$$|n, k, \alpha\rangle \equiv (\ell + 1)^{-1/2}\sum_{m=0}^{\ell} e^{-im\theta_{n,k}}\hat{F}_0^{-m}|\alpha, k\rangle, . \quad (8)$$

For clarity, we can write out these orbits as

$$|n, k, A\rangle = (\ell + 1)^{-1/2} \left\{ |A, k\rangle + e^{i\theta_{n,k}} |\delta = 0, k\rangle + \sum_{m=1}^{\ell-1} e^{im(k - \theta_{n,k})} |\delta = 2m, k\rangle \right\} \tag{9a}$$

$$|n, k, B\rangle = (\ell + 1)^{-1/2} \left\{ |B, k\rangle + e^{-i\theta_{n,k}} |\delta = 1, k\rangle + e^{-i\theta_{n,k}} \sum_{m=1}^{\ell-1} e^{im(k - \theta_{n,k})} |\delta = 2m + 1, k\rangle \right\}, \tag{9b}$$

and note that a different choice of "reference configuration" will result in an overall factor of $e^{i\theta_{n,k}}$ to some power compared to the above.

We can massage (9) into a form that looks like plane waves by rotating the $A/B$ eigenstates into symmetric or anti-symmetric linear combinations, parametrized by $\eta = \pm 1$ (unrelated to $\nu = +/- = R/L$):

$$|n, k, \eta\rangle = \frac{1}{\sqrt{2}} \left( |n, k, A\rangle + \eta e^{i\psi} |n, k, B\rangle \right), , \tag{10}$$

and consideration of the action of the Hamiltonian in section II B dictates that $\psi = (k + \theta_{n,k})/2 = k_+$. Expanding this we have

$$|k_+, k_-\rangle \propto \sum_{x=1}^{L} e^{i(k_+ + k_-)x} \left( |A, x\rangle + e^{ik_+} |B, x\rangle \right) + \sum_{\substack{x_\pm = 1 \\ x_+ \ne x_-}}^{L} e^{ik_+ x_+} e^{ik_- x_-} |x_+, x_-\rangle + e^{i(k_+ - k_-)} \sum_{x=1}^{L} e^{i(k_+ + k_-)x} |x, x\rangle, \tag{11}$$

where $\eta$ from (10) has been eliminated in favor of extending the allowed values of $(k_+, k_-)$ from the quantization of $k$ and $\theta_{n,k}$ to include $\pi$-shifted pairs $(k_+ + \pi, k_- + \pi)$. Although these states will be degenerate under $\hat{F}_0$, this distinction is necessary for the counting of states, and $\hat{H}$ will lift these degeneracies in II B.

Recalling that the molecules also correspond to particular configurations of the two movers, the coefficient of the A molecule terms are also of the same plane wave form $e^{ik_+ x_+} e^{ik_- x_-}$ as most of the doublon terms; however, the B-molecule at cell $n$ now corresponds to a right-mover at $n$ and a left-mover at $n+1$, so its coefficient is in fact $e^{i(k_+ - k_-)} e^{ik_+ x_+} e^{ik_- x_-}$, and we note that the same extra factor of $e^{i(k_+ - k_-)}$ has been applied to the same-cell doublon terms $|x, x\rangle$. These states are the respective "delay" states (i.e. the movers are in the same positions as in the preceding state under FFA) in the orbits defined in (3) and (4), respectively. Thus, we have identified the S matrix for FFA:

$$\tilde{\mathcal{S}}(k_+, k_-) = +e^{i(k_+ - k_-)}, \tag{12}$$

where, compared to the S matrices of other known integrable systems, here we have an overall sign of $+1$ rather than $-1$ due to the *distinguishability* of the two particles. Because these particles are distinguishable, we ascribe no meaning to swapping the order of the momentum arguments.

Note the fact that this S matrix multiplies only two of the configurations in (11) is an artifact of our choice of reference configuration (9). In general, the S matrix (12) appears on all post-collision configurations until the end of the "orbit", which for our choice was rather immediate. One could also make the natural choice that all terms with $x_- \le x_+$ (and the B molecule) get an S matrix, and all others do not. The quantization condition on $\theta$ ensures that there is no mismatch.

To complete our treatment of this sector, when the number of unit cells, $L = 2\ell + 1$, is *odd*, by analogy to (5) we have

$$\hat{F}_0 |m, k\rangle = e^{-i\theta_m} |m, k\rangle, \quad \theta_m = \frac{2\pi m}{L + 2} \cong \frac{2\pi m' - Lk}{L + 2}, \tag{13}$$

where here $\theta$ need not depend on $k$ since $\hat{F}_0^{L+2} = \mathbb{1}$, and compared to (5) we allow twice as many values of the integer $m$. We construct eigenstates as before as orbits under $\hat{F}_0$ starting from the $A$ molecule states for concreteness, and again recover a state of the general form (10), with the explicit value of $\psi = \pi (m + Lk/2\pi) + (k + \theta_m)/2$, where $m$ is the index of $\theta_m$. Unlike the (10), we have a single sector of size $L + 2$, and here there is no parameter $\eta$, and the overall sign of the $B$ terms relative the $A$ terms is set by $\psi$. The placement of the S matrices is the same as for even $L$ eigenstates.

In sectors with *arbitrary* numbers of particles, the eigenstates of $\hat{F}_0$ continue to be orbits constructed in a similar fashion, and based on the results for small sectors, we can guess the pattern for the quantization, which we have

confirmed numerically. In general, one has

$$K = \sum_\pm K_\pm = \frac{2\pi N_k}{L} \qquad N_k \in \{1, 2, \ldots, L-1, L\} \qquad (14a)$$

$$\Theta = \sum_\pm \pm K_\pm = \frac{2\pi N_\theta + (N_+ - N_- - L)K}{L + N_+ + N_-} \qquad N_\theta \in \{1, \ldots, L + N_+ + N_-\}, \qquad (14b)$$

provided that $L + N_+ + N_-$ is *odd*. If $L + N_+ + N_-$ is even, then we have

$$\Theta = \sum_\pm \pm K_\pm = \frac{4\pi N_\theta + (N_+ - N_- - L)K}{L + N_+ + N_-} \qquad N_\theta \in \left\{1, \ldots, \frac{1}{2}(L + N_+ + N_-)\right\}, \qquad (14c)$$

which comes from operator identities of the form $\hat{F}_0^{[\frac{1}{2}](L+N_++N_-)} \hat{T}^m = \mathbb{1}$ (for some $m$). These equations also fully determine the total momentum of all right-movers, $K_+$, and the total momentum of all left-movers, $K_-$. However, without a Hamiltonian term, it is not clear that there is a means to extract the allowed momenta of the *individual* movers, or re-write linear combinations of the various degenerate orbits as plane-waves in general many-body sectors, as one expects for integrable systems. As for the placement of S matrices in these eigenstates, one need only choose a reference state as the "default" ordering of the movers, and for each configuration with a different order of movers (due to collisions), apply corresponding S matrices as in the $N_+ = N_- = 1$ sector.

## II. THE HAMILTONIAN PERTURBATION

To generalize FFA, we will now include a dispersing Hamiltonian term in the evolution

$$\hat{F}(\lambda) = e^{-i\lambda \hat{H}} \cdot \hat{F}_0, \qquad (15)$$

where $\hat{H}$ is a local Hamiltonian that acts on a given configuration $\sigma$ of $\pm$ particles by mapping them with unit weight to all other configurations $\sigma'$ such that exactly one of the movers has been moved by a single unit cell, while preserving the number of movers of each type $N_\pm$. In cases where there are two states corresponding to a particular configuration of movers, the Hamiltonian will map to the configuration $\sigma'$ that is closest to $\sigma$ under bare FFA, thereby preserving the phase delays of the latter.

In fact, the action of this Hamiltonian is quite straightforward, however owing to the complicated nature of defining the locations of the movers, the form of $\hat{H}$ on the physical spins will *appear* quite complicated and non-generic. Writing $\hat{H}$ as the sum over local terms $\hat{H}_n$, we have

$$\hat{H}_n = \underbrace{\hat{d}_n \hat{\sigma}^+_{n+1} \hat{\sigma}^+_{n+2} \sigma^-_{n+3} \hat{\sigma}^-_{n+4} \hat{d}_{n+5}}_{\text{doublon hopping}} + \underbrace{\hat{d}_n \hat{\sigma}^+_{n+1} \hat{\sigma}^-_{n+2} \hat{d}_{n+3}}_{\text{molecule hopping}} + \underbrace{\hat{d}_n \hat{\sigma}^+_{n+1} \hat{\sigma}^+_{n+2} \hat{u}_{n+3} \hat{d}_{n+4}}_{\text{molec.} \leftrightarrow \text{doublons}} + \text{refl.}$$

$$+ \underbrace{\hat{d}_n \hat{\sigma}^+_{n+1} \hat{\sigma}^+_{n+2} \hat{\sigma}^-_{n+3} \hat{u}_{n+4} \hat{u}_{n+5} + \text{refl.}}_{\text{doublon absorption}} + \underbrace{\hat{d}_n \hat{u}_{n+1} \hat{\sigma}^+_{n+2} \hat{u}_{n+3} \hat{u}_{n+4} + \text{refl.}}_{\text{molecule absorption}} + \underbrace{\hat{u}_n \hat{u}_{n+1} \hat{\sigma}^+_{n+2} \hat{\sigma}^-_{n+3} \hat{u}_{n+4} \hat{u}_{n+1}}_{\text{exchange}} + h.c., \qquad (16)$$

where '+ refl.' indicates that one should also include the same term with the operators in reverse order, and the Hermitian conjugate of each term above should also be included; as before, $\hat{d}$ projects onto $\downarrow_z$, and $\hat{u}$ projects onto $\uparrow_z$.

### A. Single species sectors

In the single-particle sector, and single-species sector in general, only the first term in (16) acts non-trivially. This term takes the form $\hat{d}_n \hat{\sigma}^+_{n+1} \hat{\sigma}^+_{n+2} \sigma^-_{n+3} \hat{\sigma}^-_{n+4} \hat{d}_{n+5} + h.c.$, and hops a doublon (of either type) by one unit cell, provided no other particles are nearby, and hence adds to the purely chiral pseudo-energies of bare FFA, $\varepsilon = \pm k$ a more typical cosine dispersion[3]. For the single particle eigenstates of FFA, one has

$$\hat{F}(\lambda)|k, \pm\rangle = e^{-i\varepsilon}|k, \pm\rangle, \quad \text{with} \quad \varepsilon = \pm k + 2\lambda \cos(k), \qquad (17)$$

where the cosine term is independent of the species index, $\nu = \pm 1$, as $\hat{H}$ does not differentiate between the two.

We now consider the two-body sector; when necessary for concreteness, let us take them to be + particles (right movers). Since the particles are indistinguishable and cannot be placed in adjacent unit cells, this gives $\frac{1}{2}L(L-3)$ states $|x_1, x_2\rangle$ with $x_2 > x_1 + 1$. The action of $\hat{F}_0$ on these states is trivial: $\hat{F}_0|x_1, x_2\rangle = |x_1 \pm 1, x_2 \pm 1\rangle$, and amounts to translation. Since $[\hat{F}_0, \hat{H}] = 0$, eigenstates of $\hat{H}$ will automatically be eigenstates of $\hat{F}_0$. These eigenstates will take the standard coordinate Bethe Ansatz (CBA) form

$$|k_1, k_2\rangle \propto \sum_{x_2 > x_1 + 1} \left( e^{i(k_1 x_1 + k_2 x_2)} + \mathcal{S}(k_2, k_1) e^{i(k_1 x_2 + k_2 x_1)} \right) |x_1, x_2\rangle, \tag{18}$$

where $\mathcal{S}(k_2, k_1)$ is the same-species S matrix corresponding to the swapping of momenta $k_1$ and $k_2$. Unlike the S matrix for FFA that applies to particles of opposite species, the order of the momenta are important here, however the form of $\mathcal{S}$ is the same for both $\nu = \pm 1$.

The corresponding wave function is given – up to an overall normalization constant – by

$$\Psi_{k_1, k_2}(x_1, x_2) \equiv \langle x_1, x_2 | k_1, k_2 \rangle \propto \left( e^{i(k_1 x_1 + k_2 x_2)} + \mathcal{S}(k_2, k_1) e^{i(k_1 x_2 + k_2 x_1)} \right). \tag{19}$$

From this follows the "Schrödinger equation"

$$e^{-i\varepsilon_\pm(k_1, k_2)} \langle x_1, x_2 | k_1, k_2 \rangle = \langle x_1, x_2 | e^{-i\lambda \hat{H}} \hat{F}_0 | k_1, k_2 \rangle = e^{\mp i(k_1 + k_2)} \langle x_1, x_2 | e^{-i\lambda \hat{H}} | k_1, k_2 \rangle. \tag{20}$$

Because this system is integrable, one expects $\varepsilon_\pm(k_1, k_2) = \varepsilon_\pm(k_1) + \varepsilon_\pm(k_2) = \pm(k_1 + k_2) + 2\lambda \cos(k_1) + 2\lambda \cos(k_2)$, and factoring this out, one has $e^{-i\lambda E(k_1, k_2)} \langle x_1, x_2 | k_1, k_2 \rangle = \langle x_1, x_2 | e^{-i\lambda \hat{H}} | k_1, k_2 \rangle$, where $E(k_1, k_2) = 2\cos(k_1) + 2\cos(k_2)$. However, clearly $|k_1, k_2\rangle$ is an eigenstate of $\hat{H}$, independent of $\lambda$, and we can write the preceding equalities in the more familiar form

$$E(k_1, k_2) \Psi_{k_1, k_2}(x_1, x_2) = \Psi_{k_1, k_2}(x_1 - 1, x_2) + \Psi_{k_1, k_2}(x_1 + 1, x_2) + \Psi_{k_1, k_2}(x_1, x_2 - 1) + \Psi_{k_1, k_2}(x_1, x_2 + 1), \tag{21}$$

with

$$E(k_1, k_2) = 2\cos(k_1) + 2\cos(k_2) = E_1(k_1) + E_1(k_2). \tag{22}$$

The Ansatz (18) already satisfies (21) when $x_2 - x_1 > 2$ for any choice of $\mathcal{S}(k_2, k_1)$. The form of the latter factor may be determined in the usual fashion by ensuring that (21) holds – with the same eigenvalue $E(k_1, k_2)$ (22) – when $x_2 = x_1 + 2$, which gives

$$\mathcal{S}(k_2, k_1) = -e^{i(k_2 - k_1)}, \tag{23}$$

which resembles the FFA S matrix $\tilde{\mathcal{S}}$, excepting the overall factor of $-1$ and meaning of the order of the momentum arguments, both of which derive from indistinguishability. We finish our solution of the two body problem by figuring out the quantization of the momenta $k_1$ and $k_2$. Their sum $K = k_1 + k_2$ is constrained by translation invariance, an generally a quantity we will fix by hand. Additionally, one has the Bethe Ansatz Equations (BAE), obtained by bringing one of the particles around the system, i.e. demanding $\Psi_{k_1, k_2}(x_1, x_2) = \Psi_{k_1, k_2}(x_2, x_1 + L)$

$$e^{-ik_1 L} = \mathcal{S}(k_2, k_1) = e^{ik_2 L}, \tag{24}$$

although because $e^{i(k_1 + k_2)L} = 1$, this is only truly a single relation. As in other integrable models, we also have

$$\mathcal{S}(k_2, k_1) = \mathcal{S}(k_1, k_2)^* = \mathcal{S}(k_1, k_2)^{-1}. \tag{25}$$

These BAE admit straightforward solutions

$$k_2 = K - k_1, \quad k_1 = \frac{(2m+1)\pi}{L-2} - \frac{K}{L-2}, \tag{26}$$

for $m \in \{1, 2, \ldots, L-2\}$, which we have confirmed with exact numerical diagonalization.

With any number of movers of the same type, the wave functions take the form

$$\Psi_{k_1, \ldots, k_{N_\nu}}(x_1, \ldots, x_{N_\nu}) = \sum_{\text{perm}} A_{j_1, \ldots, j_{N_\nu}} \exp\left(ik_{j_1} x_1 + \cdots + ik_{j_{N_\nu}} x_{N_\nu}\right), \tag{27}$$

where a given permutation that exchanges the momenta $k_n$ and $k_{m>n}$ is accompanied by an S matrix $\mathcal{S}(k_m, k_n)$ as usual. For example, for *three* particles of the same species, $\nu$, one has wave functions

$$\Psi_{k_1,k_2,k_3}(x_1, x_2, x_3) = e^{ik_1 x_1} e^{ik_2 x_2} e^{ik_3 x_3} + \mathcal{S}_{21} e^{ik_2 x_1} e^{ik_1 x_2} e^{ik_3 x_3} + \mathcal{S}_{21} \mathcal{S}_{31} e^{ik_2 x_1} e^{ik_3 x_2} e^{ik_1 x_3}$$
$$+ \mathcal{S}_{21} \mathcal{S}_{31} \mathcal{S}_{32} e^{ik_3 x_1} e^{ik_2 x_2} e^{ik_1 x_3} + \mathcal{S}_{31} \mathcal{S}_{32} e^{ik_3 x_1} e^{ik_1 x_2} e^{ik_2 x_3} + \mathcal{S}_{32} e^{ik_1 x_1} e^{ik_3 x_2} e^{ik_2 x_3}, \qquad (28)$$

where $\mathcal{S}_{32}$ is a temporary shorthand for $\mathcal{S}(k_3, k_2)$. As more particles of the same species are added, one includes additional permuted terms to the wave function, accompanied by S matrices for the corresponding permuted momenta.

These wave functions correspond to eigenstates of both $\hat{F}_0$ and $\hat{H}$, with eigenvalue under the latter

$$\hat{H}|k_1, \ldots, k_{N_\nu}\rangle = E(k_1, \ldots, k_{N_\nu})|k_1, \ldots, k_{N_\nu}\rangle = \sum_{m=1}^{N_\nu} E_1(k_m)|k_1, \ldots, k_{N_\nu}\rangle \qquad (29)$$

and as for the two-body sector, the quantization condition obtains by bringing one particle around the system:

$$e^{ik_m L} = \prod_{\substack{n=1 \\ n \neq m}}^{N_\nu} \mathcal{S}(k_m, k_n), \qquad (30)$$

which also reduces to (24) for $N_\nu = 2$. As in that case, one also has the quantization of total momentum $\sum_m k_m = K = 2\pi q/L$ for $q \in \{1, 2, \ldots, L\}$. Together, these equations admit *exact solutions*, in contrast to most known integrable models, which we can see by noting

$$e^{ik_j L} = \prod_{\substack{j'=1 \\ j' \neq j}}^{N_\nu} S_{\nu\nu}(k_j, k_{j'}) = -\prod_{j'=1}^{N_\nu}\left(-e^{i(k_j - k_{j'})}\right) = e^{i(\pi(N_\nu - 1) + N_\nu k_j)} \prod_{j'=1}^{N_\nu} e^{-ik_{j'}} = e^{i(\pi(N_\nu - 1) + N_\nu k_j)} e^{-iK}, \qquad (31)$$

which implies $1 = e^{-i(\pi(N_\nu - 1) - K + (L - N_\nu)k_j)} = e^{2\pi i m_j}$, from which we extract the solutions for $N_\nu - 1$ of the momenta

$$k_j = \frac{1}{L - N_\nu} \left(\pi(2m_j + N_\nu - 1) - K\right), \qquad (32)$$

where $m_j \in \{1, 2, \ldots, L - N_\nu\}$, and the final momentum given by $k_{N_\nu} = K - \sum_{m=1}^{N_\nu - 1} k_m$. Additionally, one has the constraint that no two particles can have the same momentum. Lastly, (32) reduces to (26) for $N_\nu = 2$.

### B. Sectors with both species

We now consider the full Floquet drive $\hat{F}(\lambda)$ in sectors with both ± particles. The smallest such sector has $N_\pm = 1$, and we already found eigenstates for this sector in the context of $\hat{F}_0$. Re-visiting this sector will help explain many of the terms in $\hat{H}$ beyond the doublon hopping term, which was the only non-trivial term in the single-species sectors discussed in section II A.

As pointed out in section I, there are two types of collision processes under $\hat{F}_0$, corresponding to even or odd separations (or $A$ or $B$ molecules, respectively). In (3) and (4) we showed $\hat{F}_0$ cycles between states in these orbits, and we have reproduced these below, with blue arrows indicating the action of $\hat{F}_0$, black arrows indicating the action of the doublon hopping term used in the previous section II A, and red arrows indicating the action of the other Hamiltonian terms:

$$\begin{array}{c} \text{even}: \\ \\ \text{odd}: \end{array} \quad \boxed{\delta = 4} \begin{array}{c} \to \\ \updownarrow \quad \nearrow \\ \boxed{\delta = 3} \end{array} \begin{array}{c} \boxed{\delta = 2} \\ \updownarrow \quad \nearrow \\ \to \end{array} \begin{array}{c} \to \\ \boxed{\delta = 1} \end{array} \begin{array}{c} \boxed{\delta = A} \\ \updownarrow \quad \nearrow \\ \to \end{array} \begin{array}{c} \to \\ \boxed{\delta = B} \end{array} \begin{array}{c} \boxed{\delta = 0} \\ \updownarrow \quad \nearrow \\ \to \end{array} \begin{array}{c} \to \\ \boxed{\delta = L-1} \end{array} \begin{array}{c} \boxed{\delta = L-2} \\ \updownarrow \\ \to \end{array} \quad \boxed{\delta = L-3} \quad \ldots\,, \qquad (33)$$

The full Hamiltonian acts as follows on position basis states $|x_+, x_-\rangle$, $|A, x\rangle$, $|B, x\rangle$; starting with the "general" case

(the black arrows in (33)), and then listing "exceptions" (red arrows in (33)), we have

$$\hat{H}|x_+, x_-\rangle = |x_+ + 1, x_-\rangle + |x_+ - 1, x_-\rangle + |x_+, x_- + 1\rangle + |x_+, x_- - 1\rangle \tag{34a}$$

$$\hat{H}|x, x+1\rangle = |A, x\rangle + |A, x+1\rangle + |x-1, x+1\rangle + |x, x+2\rangle \tag{34b}$$

$$\hat{H}|A, x\rangle = |x, x+1\rangle + |x-1, x\rangle + |B, x\rangle + |B, x-1\rangle \tag{34c}$$

$$\hat{H}|B, x\rangle = |A, x\rangle + |A, x+1\rangle + |x, x\rangle + |x+1, x+1\rangle \tag{34d}$$

$$\hat{H}|x, x\rangle = |B, x\rangle + |B, x-1\rangle + |x+1, x\rangle + |x, x-1\rangle \tag{34e}$$

$$\hat{H}|x+1, x\rangle = |x, x\rangle + |x+1, x+1\rangle + |x+2, x\rangle + |x+1, x-1\rangle, \tag{34f}$$

where if not specifically listed above, the action of $\hat{H}$ on a state defaults to (34a).

To see that $\hat{H}$ and $\hat{F}_0$ have simultaneous eigenstates, consider $N_+ = N_- = 1$ with $L = 2\ell$ is even, in which case the eigenstates of $\hat{F}_0$ are $|n, k, \alpha\rangle$ for $\alpha = A, B$, corresponding to the type of molecule that obtains (or even/odd separations, respectively). We rotate to form eigenstates as symmetric or anti-symmetric combinations of these two,

$$|n, k, \eta\rangle = \frac{1}{\sqrt{2}} \left( |n, k, A\rangle + \eta e^{i\psi} |n, k, B\rangle \right), \tag{10}$$

with $\eta = \pm 1$ (unrelated to the species index $\nu = \pm 1$). Using equations (34), we find that

$$\hat{H}|n, k, A\rangle = 4 \cos\left(\frac{k}{2}\right) \cos\left(\frac{\theta_{n,k}}{2}\right) e^{i(k+\theta_{n,k})/2} |n, k, B\rangle, \tag{35}$$

explaining why we chose $\psi = (k + \theta_{n,k})/2 = k_+$ in section I. We also note that

$$4 \cos\left(\frac{k}{2}\right) \cos\left(\frac{\theta_{n,k}}{2}\right) = 2\cos(k_L) + 2\cos(k_R) = E(k_L, k_R) = E_1(k_L) + E_1(k_R), \tag{36}$$

as expected, apart from the factor of $\eta$ multiplying the contents of (36). We absorb the overall sign $\eta = \pm 1$ into the definition of the allowed momenta $k_+$ and $k_-$ by noting that $\cos(k \pm \pi) = -\cos(k)$, and that shifting both $k_+$ and $k_-$ by $\pi$ preserves both $k = k_+ + k_-$ and $\theta = k_+ - k_-$ modulo $2\pi$. Finally, when $L$ is odd, these two degenerate orbits merge into one, eliminating the free parameter $\eta$, which is replaced by the factor $(-1)^{m+Lk/2\pi}$ where $m$ indexes the allowed eigenvalues $\theta$ (13). It is also worth noting that $\hat{H}$ lifts degeneracies of $\hat{F}_0$, such as that of the sector with $N_\pm = 1$ with $L$ even; because of this, the full, dispersing model has spectral properties more reminiscent of conventional integrable systems.

As we add more movers of either type, not much changes compared to the above, with a few notable exceptions. The first is that new collision processes emerge: for example with $N_+ = 1$ and $N_- = 2$, $\hat{F}_0$ factorizes into its action on an isolated left-mover, and its action as in the $N_\pm = 1$ sector, with the singular exception that $\hat{F}_0|B, x\rangle \otimes |x_- = x + 2\rangle = |A, x+1\rangle \otimes |x_- = x\rangle$. However, no knowledge of these new processes is necessary to form eigenstates, since the eigenstates of $\hat{H}$ will also be eigenstates of $\hat{F}_0$. Second, we have the possibility of "cluster states" with three or more physical spins up. Looking at the same $N_+ = 1$ and $N_- = 2$ sector, $\hat{F}_0|A, x\rangle \otimes |x_- = x + 2\rangle = |x_+ = x; x_- = x, x+2\rangle$, where the latter looks like $\ldots \downarrow \uparrow_{2x-2} \uparrow_{2x-1} \uparrow_{2x} \uparrow_{2x+1} \downarrow \ldots$. For each additional mover (of alternating type), we have the possibility of a cluster with one additional up spin. In the $N_+ = 1$ and $N_- = 2$ sector, the two left-movers can only be in adjacent unit cells if the right-mover is in the same position as the first left-mover. Therefore, the right-mover is stuck: it cannot be moved by $\hat{H}$ without moving one of the left-movers at the same time. Since $\hat{H}$ can only act by moving one quasiparticle, only the outer left-movers can be moved, and they must be moved out of the cluster. This property holds for larger, generic clusters as well.

In fact, the action of $\hat{H}$ on these "cluster" configurations explains all the remaining terms not accounted for by (33). In general, all of the doublons in the cluster are locked in place, and $\hat{H}$ can only act by bringing movers into or out of the cluster at the edges, either in the form of molecules or two-body clusters ($\uparrow\uparrow\uparrow$). Lastly, there is a term allowing for one mover to hop between adjacent clusters. It is also worth noting that the same-species S matrix $\mathcal{S}(k_2, k_1)$ is designed to give zero weight to a state with movers of the same type on adjacent sites. One might then worry that these cluster configurations will have zero weight; fortunately, these states are also "delay states", i.e. two (or more) of the movers will be in the same place if one acts with $\hat{F}_0^{-1}$, and therefore they will also have a factor of $\tilde{\mathcal{S}}$ that will prevent the weight on these states from vanishing.

## C. Form of the solution

For generic many-body sectors, the eigenstates are constructed from plane waves (solutions to the single-particle sector) as in Hamiltonian integrable systems. First, we write the "naïve" wave function for a given configuration as the product of the wave functions within the two sectors (27), i.e. $\Psi_{k_1^+,\ldots,k_{N_+}^+;k_1^-,\ldots,k_{N_-}^-}(x_1^+,\ldots,x_{N_+}^+;x_1^-,\ldots,x_{N_-}^-) = \Psi_{k_1^+,\ldots,k_{N_+}^+}(x_1^+,\ldots,x_{N_+}^+)\Psi_{k_1^-,\ldots,k_{N_-}^-}(x_1^-,\ldots,x_{N_-}^-)$, each of which is the sum over permutations of the momenta in that sector. Unlike the single-species sectors, we will also have molecule configurations, for which the form of the wavefunction is identical to the doublon configurations: plane waves evaluated at the prescribed positions of the molecules' constituent movers. We then identify a "reference" configuration, i.e. an ordering of all the ± movers, starting from the first unit cell, and compared to this configuration, any configuration in which the order of a + and − is interchanged picks up an S matrix for that collision, $\tilde{\mathcal{S}}(k_n^+, k_m^-)$.

This is most easily seen in sectors with sufficiently low density to admit a state where all − particles are to the right of all + particles; in this case, every state with the $m^{\text{th}}$ − left of the $n^{\text{th}}$ + will be multiplied by $\tilde{\mathcal{S}}(k_n^+, k_m^-)$. Since the wave function is a superposition over different assignments of the momenta $\{k_n^+, k_m^-\}$ to the various movers, each individual term in the sum will have a different momentum appearing in any particular factor of $\tilde{\mathcal{S}}$, as $\tilde{\mathcal{S}}$ is associated to the movers. In summary:

$$\Psi_{k_1^+,\ldots,k_{N_+}^+;k_1^-,\ldots,k_{N_-}^-}(x_1^+,\ldots,x_{N_+}^+;x_1^-,\ldots,x_{N_-}^-) =$$
$$\sum_{\text{perm}} A^+_{n_1,\ldots,n_{N_+}} A^-_{m_1,\ldots,m_{N_-}} \varsigma\left(k_{n_1}^+, \ldots k_{n_{N_+}}^+; k_{m_1}^-, \ldots, k_{m_{N_-}}^-\right) \exp\left(ik_{n_1}^+ x_1^+ + ik_{m_1}^- x_1^- + \cdots + ik_{n_{N_+}}^+ x_{N_+}^+ + ik_{m_{N_-}}^- x_{N_-}^-\right), \quad (37)$$

where the coefficients $A$ are unity for unpermuted labels, and acquire factors $\mathcal{S}_{21}$ if the momenta of the first and second mover (of a given species) are swapped. The factor $\varsigma$ is a place-holder for the product of necessary +− S matrices, $\tilde{\mathcal{S}}(k_{n_j}^+, k_{m_i}^-)$, relative some "default" configuration of the movers.

The structure of these eigenstates gives rise to the quantization of the momenta. Recall that the quantization of $\hat{F}_0$, which is independent of $\hat{H}$, constrains $\Theta = K_+ - K_- = \sum_\pm \pm K_\pm$, and as always, translation invariance dictates the quantization of total momentum $K = K_+ + K_- = \sum_\pm K_\pm$, per (14). Even in a trivial limit wherein we omitted $\hat{F}_0$ from our Floquet drive, the fact that $[\hat{F}_0, \hat{H}] = 0$ would still lead to the same quantization condition. Since a quantization condition exists for both the sums and differences of $K_\pm$, the two are both independently fixed. What remains is a formula of the form (30) for scenarios in which both movers are present. As in that case, these conditions obtain from moving one quasiparticle around the entire system, and correspondingly, contains S matrices corresponding to all of the resultant collisions. Thus, compared to (30), one expects the full quantization condition to contain S matrix factors corresponding to collisions between quasiparticles of *opposite* chirality as well. Indeed, those equations are

$$e^{ik_j^\nu L} = \prod_{\substack{n=1\\n\neq j}}^{N_\nu} \mathcal{S}\left(k_j^\nu, k_n^\nu\right) \prod_{m=1}^{N_{\bar\nu}} \tilde{\mathcal{S}}^{\bar\nu}\left(k_j^\nu, k_m^{\bar\nu}\right) \quad (38a)$$

where again, there is no meaning to switching the order of the momenta for $\tilde{\mathcal{S}}$. For right-movers with momenta $p_j$, this takes the form

$$e^{ip_j L} = \prod_{\substack{n=1\\n\neq j}}^{N_+} \mathcal{S}(p_j, p_n) \prod_{m=1}^{N_-} \tilde{\mathcal{S}}^{-1}(p_j, q_m), \quad (38b)$$

$$= -\prod_{n=1}^{N_+}\left[-e^{i(p_j - p_n)}\right] \prod_{m=1}^{N_-}\left[e^{i(q_m - p_j)}\right] \quad (38c)$$

$$= (-1)^{N_+ - 1} e^{i(N_+ p_j - K_+)} e^{i(K_- - N_- p_j)}, \quad (38d)$$

and for left-movers with momenta $q_j$, the form

$$e^{iq_j L} = \prod_{\substack{n=1\\n\neq j}}^{N_-} \mathcal{S}(q_j, q_n) \prod_{m=1}^{n_+} \tilde{\mathcal{S}}(p_m, q_j) \quad (38e)$$

$$= -\prod_{n=1}^{N_-}\left[-e^{i(q_j - q_n)}\right] \prod_{m=1}^{N_+}\left[e^{i(p_m - q_j)}\right] \quad (38f)$$

$$= (-1)^{N_- - 1} e^{i(N_- q_j - K_-)} e^{i(K_+ - N_+ q_j)}. \quad (38g)$$

In either case, one then has

$$e^{ik_j^\pm L} = (-1)^{N_\pm - 1} e^{i(N_\pm k_j^\pm - K_\pm)} e^{i(K_\mp - N_\mp k_j^\pm)} \tag{38h}$$

which implies

$$e^{2\pi i m_j^\pm} = 1 = e^{i(L - N_\pm + N_\mp)k_j^\pm} e^{i(K_\pm - K_\mp)} (-1)^{N_\pm - 1}, \tag{38i}$$

from which we can easily extract the solutions

$$k_j^+ = \frac{\pi(2m_j^+ + N_+ - 1) - \Theta}{L - N_+ + N_-} \qquad m_j^+ \in \{1, 2, \ldots, L - N_+ + N_-\} \tag{39a}$$

$$k_j^- = \frac{\pi(2m_j^- + N_- - 1) + \Theta}{L - N_- + N_+} \qquad m_j^- \in \{1, 2, \ldots, L - N_- + N_+\}. \tag{39b}$$

The solutions to this model consist of all $\{k_j^\pm\}$ of the form (39) where $K$ and $\Theta$ satisfy (14), and $\sum_{j=1}^{N_\pm} k_j^\pm = K_\pm = \frac{1}{2}(K \pm \Theta)$. Additionally, no two movers of the same type may have the same momentum. These BAE have been tested against exact numerics for small systems, and appear to hold for all accessible sizes (roughly 22 physical spins and $N_\pm \sim L/2$, and out to larger $L$ for lower filling fractions). The corresponding eigenvalues under the combined action are given by

$$e^{-i\lambda \hat{H}} \hat{F}_0 |k_1^+, \ldots, k_{N_+}^+; k_1^-, \ldots, k_{N_-}^-\rangle = e^{-i\varepsilon(k_1^+, \ldots, k_{N_+}^+; k_1^-, \ldots, k_{N_-}^-)} |k_1^+, \ldots, k_{N_+}^+; k_1^-, \ldots, k_{N_-}^-\rangle, \tag{40}$$

with

$$\varepsilon(k_1^+, \ldots, k_{N_+}^+; k_1^-, \ldots, k_{N_-}^-) = \sum_\pm \sum_{j=1}^{N_\pm} \left\{ \pm k_j^\pm + 2\lambda \cos(k_j^\pm) \right\}. \tag{41}$$

## III. THERMODYNAMICS AND HYDRODYNAMICS

Even though the BAE (38) admit exact and simple solutions (39) even as $L \to \infty$, it will still be useful to go through the standard procedure of the thermodynamic Bethe Ansatz (TBA) that has been successfully applied to other integrable models[4]. In particular, the exceptionally simple nature of this model's S matrices will translate into simple results in TBA. As well, this will give us access to statistical mechanics of this model in the so-called generalized Gibbs ensemble (GGE), which stands apart from previously studied integrable systems not only in its simplicity, but the Floquet nature of the model means we must forego the standard "free energy". We will find it convenient to introduce the following function:

$$\mathcal{S}_0(k_2 - k_1) = e^{i(k_2 - k_1)}, \tag{42}$$

such that $\mathcal{S}(k_2, k_1) = -\mathcal{S}_0(k_2 - k_1) == \mathcal{S}^*(k_1, k_2)$ for either $\nu = \pm 1$, and $\tilde{\mathcal{S}}(k_m^+, k_n^-) = \tilde{\mathcal{S}}(k_n^-, k_m^+) = +\mathcal{S}_0(k_m^+ - k_n^-)$.

### A. Bethe equations in the thermodynamic limit

We begin with quantization condition (39)

$$2\pi m_j^\nu = L k_j^\nu + i \sum_{\substack{n=1 \\ n \neq j}}^{N_\nu} \ln -\mathcal{S}_0(k_j^\nu - k_n^\nu) + i \sum_{m=1}^{N_{\bar\nu}} \ln \mathcal{S}_0(k_m^{\bar\nu} - k_j^\nu), \tag{43}$$

and replace the quantum numbers $m_j^\nu$ with 'counting numbers' $L c_\nu(k)$. When $L c_\nu(k) = m_j^\nu$, if the momentum mode corresponding to $m_j^\nu$ is occupied, one has a "particle", and if it is not, one has a "hole".

$$c_\nu(k) = \frac{k}{2\pi} + \frac{i}{2\pi L} \left\{ i\pi N_\nu + \sum_{n=1}^{N_\nu} \ln \mathcal{S}_0(k_j^\nu - k_n^\nu) - \sum_{m=1}^{N_{\bar\nu}} \ln \mathcal{S}_0(k_m^{\bar\nu} - k_j^\nu) \right\}. \tag{44}$$

Of course it is possible to further simplify (48) substantially, however we abstain from doing so to maintain consistency with other works on TBA. We obtain the total density of states (particles and holes) by differentiating the counting function with respect to its momentum arguments

$$\rho_\nu(k) + \bar\rho_\nu(k) = \rho_\nu^{\text{tot}}(k) = \frac{dc_\nu}{dk}(k), \tag{45}$$

where $\bar\rho$ is the density of holes. We will implicitly make use of the relation $\frac{1}{L}\sum_{n\neq j}(p_{(j)} - p_n) \to \int_{-\pi}^{\pi} dp'\,(p_{(j)} - p')\rho_\nu(p')$. To match with the main text and other references on integrable systems, we define a matrix version of the S matrix (in the species space)

$$S_{\nu\nu'}(q_2, q_1) = \delta_{\nu\nu'}\mathcal{S}(q) + (1 - \delta_{\nu\nu'})\tilde{\mathcal{S}}^{\nu'}(q), \tag{46}$$

where in the second term $\nu'$ is an exponent, not a superscript. We define the 'kernel'

$$\mathcal{K}_{\nu\nu'}(p, p') \equiv \frac{1}{2\pi i}\frac{d}{dp}\ln S_{\nu\nu'}(p - p') = \frac{1}{2\pi i}(i\delta_{\nu\nu'} - i(1 - \delta_{\nu\nu'})) = \frac{1}{2\pi}(2\delta_{\nu\nu'} - 1) = \frac{\nu\nu'}{2\pi}, \tag{47}$$

when we regard $\nu, \nu' = \pm 1$. The total density of states (DoS) are therefore given by

$$\rho_\nu^{\text{tot}}(p) = \frac{1}{2\pi}\left\{1 - \int_{-\pi}^{\pi} dp'\,\rho_\nu(p') + \int_{-\pi}^{\pi} dq\,\rho_{\bar\nu}(q)\right\} = \frac{1}{2\pi}\left(1 + \frac{N_\nu - N_{\bar\nu}}{L}\right) = \frac{1}{2\pi}(1 - n_\nu + n_{\bar\nu}), \tag{48a}$$

where $n_\nu = \frac{N_\nu}{L} = \int_{-\pi}^{\pi} dk\,\rho_\nu(k)$. Summarizing these results:

$$\rho_\nu(k) + \bar\rho_\nu(k) = \rho_\nu^{\text{tot}}(k) = \frac{dc_\nu}{dk}(k) = \frac{1}{2\pi}(1 + n_{\bar\nu} - n_\nu) = \frac{1}{2\pi}\left\{1 + \int_{-\pi}^{\pi} dk'\,[\rho_{\bar\nu}(k') - \rho_\nu(k')]\right\}, \tag{49}$$

which means that $\rho_\nu$ is independent of $k$.

### B. Effective Thermodynamic Bethe Ansatz (TBA) partition function

We now construct an effective partition function with respect to which one can compute thermodynamic expectation values, etc. in the model's generalized Gibbs ensemble (GGE). The partition function is the exponential of an effective "free energy", which is a poor choice of name for a Floquet system, and so we will call it the "[Generalized] Gibbs Potential" (GGP), or symbolically: $G = L\mathcal{G}$. Because the model is Floquet, we do not define a temperature $T$, and correspondingly, we do not insist other quantities inherit units of energy:

$$\mathcal{Z} = e^{-G} = \int \mathcal{D}[\rho_+, \rho_-]\,e^{-L\mathcal{G}[\rho_+, \rho_-]} \tag{50}$$

$$= \int \mathcal{D}[\rho_+, \rho_-]\,e^{-\mu_+ N_+}\,e^{-\mu_- N_-}\,e^{L\,S_{YY}[\rho_+, \rho_-]}, \tag{51}$$

where $S_{YY}$ is the Yang-Yang entropy function corresponding to the densities of states $\rho_\nu$. For concreteness, we restricted ourselves to a GGE specified by two chemical potentials $\mu_\pm$ for the two-quasiparticle species. The GGP reads

$$\mathcal{G}[\rho_+, \rho_-] = \mu_+ n_+ + \mu_- n_- - S_{YY}[\rho_R, \rho_-] \tag{52}$$

$$= \sum_{\nu=\pm 1}\int_{-\pi}^{\pi} dk\,\left\{\mu_\nu \rho_\nu - [\rho_\nu^{\text{tot}}\ln\rho_\nu^{\text{tot}} - \rho_\nu \ln\rho_\nu - \bar\rho_\nu \ln\bar\rho_\nu]\right\}. \tag{53}$$

Since the partition function is of the form $\mathcal{Z} = \int \mathcal{D}\rho\,e^{-L\mathcal{G}[\rho]}$ as $L \to \infty$, we can compute this integral by saddle point, and so we must find extrema of $\mathcal{G}$. Thus, we will use a functional variation, sending $\rho_\nu \to \rho_\nu + \delta\rho_\nu$, and correspondingly

$\mathcal{G} \to \mathcal{G} + \delta\mathcal{G}$, and choose $\rho_\nu$ such that $\delta\mathcal{G} \to 0$. We will also make use of the fact that $\rho$ and $\bar{\rho}$ are related by the Bethe equation (49), and derive a variational relation therefrom:

$$\delta\rho_\nu^{\text{tot}}(k) = \delta\rho_\nu(k) + \delta\bar{\rho}_\nu(k) = \frac{1}{2\pi}\int_{-\pi}^{\pi} dk' \left[\delta\rho_{\bar{\nu}}(k') - \delta\rho_\nu(k')\right] \tag{54}$$

and with this, we are ready to extremize the GGP:

$$\delta\mathcal{G} = \sum_{\nu=\pm 1}\int_{-\pi}^{\pi} dk \left\{\mu_\nu \delta\rho_\nu - \left[\delta\rho_\nu \ln\frac{\rho_\nu^{\text{tot}}}{\rho_\nu} + \delta\bar{\rho}_\nu \ln\frac{\rho_\nu^{\text{tot}}}{\bar{\rho}_\nu}\right]\right\}, \tag{55}$$

where so far, all $\rho$'s have been functions of $k$, but now we invoke (54) and must be explicit about this. After a few standard manipulations[4] one has

$$\delta\mathcal{G} = \sum_{\nu=\pm 1}\int_{-\pi}^{\pi} dk\, \delta\rho_\nu(k) \left\{\mu_\nu - \left[\ln\frac{\bar{\rho}_\nu}{\rho_\nu} - \int_{-\pi}^{\pi} \frac{dq}{2\pi}\ln\left(1 + \frac{\rho_\nu(q)}{\bar{\rho}_\nu(q)}\right) + \int_{-\pi}^{\pi}\frac{dq}{2\pi}\ln\left(1 + \frac{\rho_{\bar{\nu}}(q)}{\bar{\rho}_{\bar{\nu}}(q)}\right)\right]\right\}, \tag{56}$$

which is zero if the quantity in curly braces above is zero. Writing $\frac{\bar{\rho}_\nu(k)}{\rho_\nu(k)} = e^{\epsilon_\nu(k)}$, the extremization is guaranteed by the TBA equations:

$$\epsilon_\nu(k) = \mu_\nu + \int_{-\pi}^{\pi}\frac{dq}{2\pi}\ln\frac{1 + e^{-\epsilon_\nu(q)}}{1 + e^{-\epsilon_{\bar{\nu}}(q)}}. \tag{57}$$

These equations imply that $\epsilon_\nu$ does not depend on momentum, so that

$$\epsilon_\nu = \mu_\nu + \ln\frac{1 + e^{-\epsilon_\nu}}{1 + e^{-\epsilon_{\bar{\nu}}}}, \tag{58}$$

In the case where $N_+ = N_-$, whence one has $\mu_+ = \mu_- = \mu$, the equations simplify even further to $\epsilon_\nu = \mu$ for both $\nu = \pm 1$.

### C. Generalized hydrodynamics

As discussed in the main text, the generalized hydrodynamics[5,6] of the DFFA model follows immediately from the TBA equations above. Let us consider a (generalized) equilibrium state with filling $n = \frac{1}{1+e^\mu}$, corresponding to $\mu_+ = \mu_- = \mu$. The effective (dressed) velocities of the quasiparticles in this simple GGE follow from the equation given in the main text, and they are given by

$$v_{\pm,k} = \pm\frac{1}{1 + 2n} - 2\lambda \sin k. \tag{59}$$

As in generic interacting integrable models, the quasiparticle trajectories broaden diffusively due to the random collisions with other quasiparticles. The variance of the position of a given quasiparticle at time $t$ is given by[7,8]

$$\delta x_{\nu,k}^2(t) = t\frac{1}{(\rho_{\nu,k}^{\text{tot}})^2}\sum_{\nu'}\int dk' |v_{\nu,k} - v_{\nu',k'}|\left[\mathcal{K}_{\nu\nu'}^{\text{dr}}(k,k')\right]^2 \rho_{\nu',k'}(1 - \theta_{\nu',k'}), \tag{60}$$

where

$$\mathcal{K}_{\nu\nu'}^{\text{dr}}(k,k') = \mathcal{K}_{\nu\nu'}(k,k') - \sum_{\nu''}\int dk'' \mathcal{K}_{\nu\nu''}(k,k'')\mathcal{K}_{\nu''\nu'}^{\text{dr}}(k'',k')\theta_{\nu'',k''}, \tag{61}$$

is the "dressed kernel", and $\theta_{\nu,k} = \frac{\rho_\nu}{\rho_\nu^{\text{tot}}}$ is the filling fraction or Fermi factor. Let us evaluate these formulas in an equilibrium state with chemical potential $\mu$, corresponding to a given filling $n$. The equations for the dressed Kernels become

$$\mathcal{K}_{++}^{\text{dr}} = \frac{1}{2\pi} - n\left(\mathcal{K}_{++}^{\text{dr}} - \mathcal{K}_{+-}^{\text{dr}}\right), \tag{62}$$

$$\mathcal{K}_{+-}^{\text{dr}} = -\frac{1}{2\pi} - n\left(\mathcal{K}_{+-}^{\text{dr}} - \mathcal{K}_{++}^{\text{dr}}\right), \tag{63}$$

so that

$$\mathcal{K}^{\text{dr}}_{++} = \mathcal{K}^{\text{dr}}_{--} = \frac{1}{2\pi(1+2n)}, \quad \mathcal{K}^{\text{dr}}_{+-} = \mathcal{K}^{\text{dr}}_{-+} = -\frac{1}{2\pi(1+2n)}. \tag{64}$$

These explicit expressions for the dressed kernels combined with the dressed velocities (59) fully determine the quasi-particle broadening using eq. (60). This expression simplifies even further in the case of the pure FFA model ($\lambda = 0$). Focusing on a right mover, the effective velocity reads $\pm v_\pm = 1 - \frac{2n}{1+2n} = \frac{1}{1+2n}$. This yields

$$\delta x^2(t) = t(2\pi)^2 \left( \int_{-\pi}^{\pi} dk' \right) |2v_+| (\mathcal{K}^{\text{dr}}_{+-})^2 \frac{1}{2\pi} n(1-n) = t \frac{2n(1-n)}{(1+2n)^3}. \tag{65}$$

This coincides with the formula found in Ref. 8 using a more elementary transfer matrix approach.

---



\end{document}